\documentclass[preprint,proceedings]{rmaa}

\suppressfulladdresses 

\usepackage{paralist}

\usepackage{psfrag,color}

\SetYear{2007}
\SetConfTitle{Massive Stars: Fundam. Param. and Circum. Interac. }

\title{Accurate Quantitative Spectroscopy of OB Stars: 
\\C and N abundances near the Main Sequence} 

\author{
 M. F. Nieva\altaffilmark{1,2} \& N. Przybilla\altaffilmark{1} }

\altaffiltext{1}{Dr. Remeis Sternwarte Bamberg, Sternwartstr. 7, D-96049 Bamberg, Germany
}
\altaffiltext{2}{Observat\'orio Nacional, Rua General Jos\'e Cristino 77 CEP 20921-400, Rio de Janeiro, Brazil}

\shortauthor{Nieva \& Przybilla}
\shorttitle{C and N abundances}

\listofauthors{Nieva, M. F. \& Przybilla, N.}
\indexauthor{Nieva, M. F.}
\indexauthor{Przybilla, N.}

\resumen{Presentamos una avanzada t\'ecnica de an\'alisis capaz de 
reproducir el espectro completo de H y He en estrellas
de tipo OB en el visible y el IR cercano y de derivar simult\'aneamente 
 abundancias met\'alicas de alta precisi\'on (hasta el momento C y N).
La s\'intesis espectral cuenta con una soluci\'on no-ETL h\'ibrida
usando nuestros modelos at\'omicos m\'as recientes.
Derivamos espectrosc\'opicamente (a partir de l\'ineas de H ensanchadas por efecto Stark y equilibrios de ionizaci\'on de He\,{\sc i/ii} y C\,{\sc ii-iv})
par\'ametros atmosf\'ericos exactos, pr\'acticamente libres de errores 
sistem\'aticos, para una muestra de estrellas distribuidas 
al azar en la vecindad solar.
Encontramos abundancias altamente consistentes, 
en contraste con previos resultados indicando amplia dispersi\'on y grandes incertezas. 
Las mejoras resultan de evitar errores sistem\'aticos en la 
determinaci\'on de par\'ametros, que pueden ser mayores que los esperados
 en trabajos anteriores, 
y en una evaluaci\'on cr\'itica de los datos at\'omicos para la construcci\'on de los modelos at\'omicos.
}

\abstract{We present a state-of-the-art analysis technique able to simultaneously
reproduce the entire H and He spectra of OB-type stars in the visual
and the near-IR and to derive highly accurate metal abundances (so far C and~N).
The spectrum synthesis relies on a hybrid non-LTE approach
involving our most recent model atoms.
Accurate atmospheric parameters, practically free of systematic
errors, are derived spectroscopically (from Stark-broadened H lines and
ionization equilibria of He\,{\sc i/ii} and C\,{\sc ii-iv}) for a sample of randomly distributed stars in the solar vicinity.
Highly consistent abundances
are found in contrast to previous reports indicating broad scatter and large uncertainties. The improvements result
from avoidance of systematic errors in the parameter determination, which may be
larger than expected in previous work, and a critical evaluation of atomic data for the model atom construction.
}

\addkeyword{Stars: early type}
\addkeyword{Stars: abundances}
\addkeyword{Line: formation}

\begin{document}
\maketitle

\section{Introduction}
\label{sec:intro}
Detailed chemical composition studies of early-type stars
contribute to the understanding of broad fields
like galactochemical and stellar evolution, 
 providing observational constraints to theory.
Therefore, the goal should be to derive chemical
 abundances as accurately as possible in order to provide
 tight constraints.
The results rely on many model assumptions,
since the quantitative analysis of observed spectra 
requires the solution of the atmospheric structure equations and  
the radiative transfer. In particular, 
the interaction of radiation and stellar plasma 
with a realistic description of the atomic processes involved 
has to be accounted for.

In a series of papers (Nieva \& Przybilla~2006ab, 2007, 2008)
we provide details of our state-of-the-art quantitative spectral analysis of 
H, He\,{\sc i/ii} and C\,{\sc ii-iv} for OB-type stars.
Here, we report, in addition, preliminary results for N
 based on the parameters derived in the
 previous analysis and an improved version of 
 the model atom  by Przybilla \& Butler~(2001).
We derive highly homogeneous abundances
for C and slightly inhomogeneous N abundances,
quantitatively consistent with predictions of stellar and galactochemical 
evolution models.

\section{Analysis \& Results}
\label{sec:models}

A hybrid approach is used for the non-LTE line formation computations. These are
based on line-blanketed plane-parallel, homogeneous and hydrostatic LTE model
atmospheres calculated with ATLAS9. Non-LTE synthetic
spectra are computed with recent versions of DETAIL and SURFACE. These codes solve the coupled radiative
transfer and statistical equilibrium equations and compute synthetic spectra
using refined line-broadening data.

A first sample of six apparently slow-rotating early-type dwarfs and 
giants from OB associations and from the field in the solar vicinity is analysed.

The stellar parameters are derived from application of an extensive iterative method
 resulting in simultaneous fits to almost all measurable H, He\,{\sc i/ii} and C\,{\sc ii-iv} lines.
 The iteration is performed on effective temperature and surface gravity  (in order to achieve ionization balance) as well as the micro-, macroturbulent and
 projected rotational velocities (from carbon line profiles) and He and C
abundances.
The final set of stellar parameters is confirmed by the spectral energy distribution fits from the UV to the near-IR.
This state-of-the-art analysis technique based on new model atoms with
critically selected atomic data and observed spectra 
of excellent quality allows us to derive highly accurate stellar parameters and chemical abundances with extremely reduced systematic errors.
A large quantity of spectral lines is analysed for the first time, giving
simultaneously consistent parameters and abundances for all of them
(H: $\sim$10 lines, He: $\sim$20, C:~$\sim$30, N: $\sim$70).

\begin{figure}[!t]
   \includegraphics[width=\columnwidth]{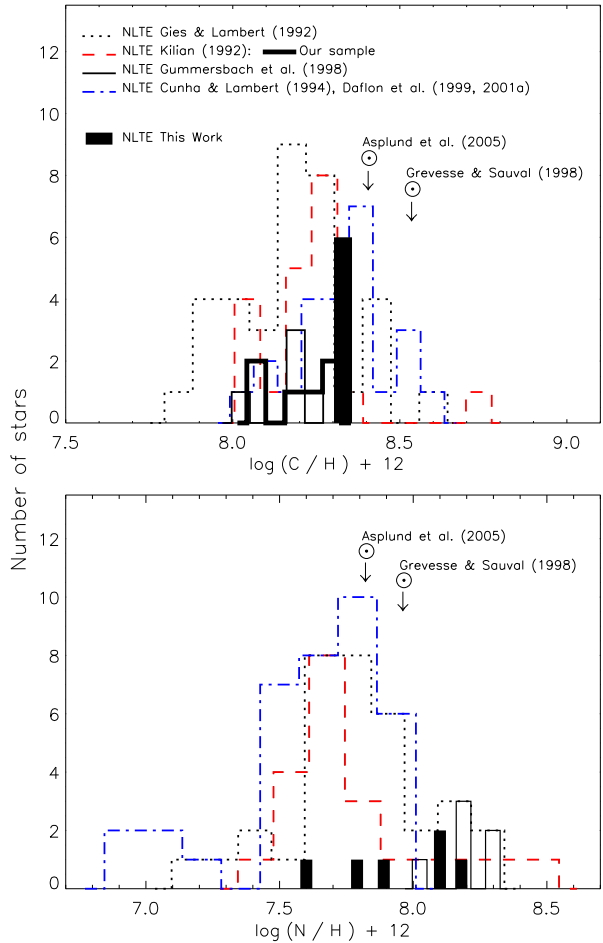}\\[-8mm]
   \caption{C and N abundances of OB dwarfs and giants in the solar vicinity
   (R$_g-$R$_g^\odot \le 500$~pc).}
   \label{fig:hist}
 \end{figure}

Figure~\ref{fig:hist} shows a comparison of our first results for present-day 
C and N abundances of early B-type stars in the solar vicinity with results from representative 
sources in the literature and with recent solar values. The stars are located 
at distances shorter than 1\,kpc from the Sun and
at galactocentric distances within up to 500\,pc difference with
respect to the location of the Sun
(see Nieva \& Przybilla 2008 for details).
Despite the small sample size, the programme stars provide
homogeneous C abundances. Kilian~(1992) performed the most
consistent spectroscopic analysis at that time, but her abundances
for our sample stars are systematically lower than ours and 
the spread is larger.
For C we derive present-day abundances that are almost solar (reference:
Asplund et al.~2005) or slightly sub-solar (reference: Grevesse \& Sauval~1998).
The derived N abundances show a slightly inhomogeneous
distribution because of rotational mixing effects in the course of stellar evolution.
This implies that the pristine C abundances (i.e. unaltered by stellar evolution)
are slightly larger 
than indicated, and the N abundances lower. Using the constraint of number
conservation in the CN cycle we can conclude that the pristine C and N
abundances of these young stars are in excellent agreement with the solar
values of Asplund et al. (a chemical peculiarity of one object cannot be
excluded at present).
The finding of a narrow distribution of C abundances 
remains basically unaffected by stellar evolution effects and it also agrees with predictions of
galactochemical evolution (e.g. Chiappini et al.~2003).
We conclude that the reduction of systematic effects in the whole 
quantitative spectral analysis leads to highly consistent values of chemical abundances. A larger sample has to be analysed in order to improve on the statistics.
 
\end{document}